\pdfoutput=1
\documentclass[aps,prl,reprint,amssymb,amsmath,superscriptaddress,showkeys]{revtex4-1}
\usepackage{lmodern}
\usepackage[latin9]{inputenc}
\usepackage{color}
\usepackage{amsmath}
\usepackage{graphicx}
\usepackage{microtype}

\makeatletter

\usepackage{dcolumn}
\usepackage{bm}
\usepackage{color}
\usepackage[makeroom]{cancel}

\definecolor{blue}{rgb}{0.13, 0.17, 0.56}

\makeatother

\begin{document}

\title{Growing Critical: Self-Organized Criticality in a Developing Neural
System}

\author{Felipe Yaroslav Kalle Kossio}

\author{Sven Goedeke}

\affiliation{Neural Network Dynamics and Computation, Institute of Genetics, University
of Bonn, Bonn, Germany}

\author{Benjamin van den Akker}

\affiliation{Department of Neuroinformatics, Radboud University Nijmegen, Nijmegen,
Netherlands}

\author{Borja Ibarz}

\affiliation{Nonlinear Dynamics and Chaos Group, Departamento de Fisica, Universidad
Rey Juan Carlos, Madrid, Spain}

\author{Raoul-Martin Memmesheimer}

\affiliation{Neural Network Dynamics and Computation, Institute of Genetics, University
of Bonn, Bonn, Germany}

\affiliation{Department of Neuroinformatics, Radboud University Nijmegen, Nijmegen,
Netherlands}

\begin{abstract}
Experiments in various neural systems found avalanches: bursts of
activity with characteristics typical for critical dynamics. A possible
explanation for their occurrence is an underlying network that self-organizes
into a critical state. We propose a simple spiking model for developing
neural networks, showing how these may ``grow into'' criticality.
Avalanches generated by our model correspond to clusters of widely
applied Hawkes processes. We analytically derive the cluster size
and duration distributions and find that they agree with those of
experimentally observed neuronal avalanches.
\end{abstract}
\maketitle
\textit{Introduction.---}A hallmark of systems at criticality is
the variability of their responses to small perturbations. While small
responses are most likely, the probability of large, system-size effects
is non-negligible. Various natural and model complex systems show
similar behavior \citep{markovic2014power}. One explanation is that
they drive themselves close to a critical state (``self-organized
criticality'' \citep{BTW87,BTW88}). The dynamics of such systems
are characterized by ``events'' or ``avalanches''. Their sizes
and durations follow power-law distributions, frequently with exponents
$3/2$ and $2$, indicating an underlying critical branching process
\citep{ZLS95,Jensen98,harris2002theory,santo2017simple}. Apparent
critical dynamics, ``neuronal avalanches'', in biological neural
networks were first reported in Refs.~\citep{BP03,BP04}. It has
been suggested that they foster information storage and transfer \citep{haldeman2005,shew2013}.
Experimental studies often report power-law size and duration distributions
with exponents $3/2$ and $2$. They further indicate that neuronal
avalanches emerge during development \citep{MBGBRT07,GP08,hennig2009,yada2017development},
suggesting that neural networks develop into a critical state. 

The development of neural networks is determined by an interplay of
genetic determinants and environmental influence. Of pivotal importance
is neural activity \citep{KM91,KMCC88}. As a general rule, neurons
with low activity level extend their neurites and form more activating
connections, while highly active cells reduce these \citep{CK86,HR87,FNN90}.
Thereby, neurons maintain their average activity at a particular level
(homeostasis) \citep{Ooyen11,turrigiano2012homeostatic,ooyen2017rewiring}.

Computational models for avalanches in neural systems rely on static,
tuned connectivity \citep{EHE02,hennig2009}, on short-term synaptic
plasticity \citep{LHG07,LHG09}, or on long-term network changes \citep{abbott2007,GKLK08,TOEWB10,delpapa2017}.
Here we propose a continuous-time spiking neural network model belonging
to the third class. The avalanche dynamics follow from a network growth
process towards a critical state, which uses local information only
\citep{BR00,BR03,MG09}. The model is rooted in previous models for
neural network development \citep{OP94,abbott2007,TOEWB10}, but sufficiently
simple to be analytically tractable.

\emph{Neuron model.---}Like biological neurons, our model neurons
communicate by sending and receiving spikes in continuous time. Spiking
is stochastic, according to an inhomogeneous Poisson point process
with instantaneous rate $f_{i}(t)$ for neuron $i$ \citep{GK06,kempter1999hebbian,abbott2007,BGH07,PSCR11}.
In isolation neurons have a low spontaneous rate $f_{0}$, e.g., due
to spontaneous synaptic release or channel fluctuations \citep{koch1999biophysics,ford2012}.
A spike from neuron $j$ increases $f_{i}$ by the size of the time-dependent
connection strength $gA_{ij}$. The increment decays exponentially
with time constant $\tau$, which accounts for relaxation due to leak
currents. The couplings are excitatory; this is the dominant connection
type in developing neural systems \citep{OP94}. Taken together, $f_{i}$'s
dynamics follow
\begin{equation}
\tau\dot{f_{i}}(t)=f_{0}-f_{i}(t)+\tau g\sum_{j}A_{ij}(t^{-})\sum_{\hat{t}_{j}}\delta\left(t-\hat{t}_{j}\right),\label{hw_model_rate}
\end{equation}
where $\hat{t}_{j}$ denotes the spike times of neuron $j$ ($\delta$
is the Dirac delta distribution). For simplicity, we assume that all
neurons have the same parameters. For constant couplings, the network
dynamics form a multivariate Hawkes point process \citep{hawkes1971,BGH07,PSCR11}.

\begin{figure}
\noindent \begin{centering}
\includegraphics[width=1\columnwidth]{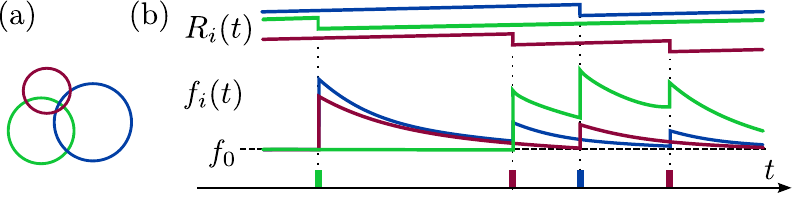}
\par\end{centering}
\caption{\label{fig:MicroscopicModelDynamics} Neuron dynamics. (a) Neurons'
somas and neurite extents are represented by disks with evolving radii.
Coupling strengths are proportional to neurite overlap areas. (b)
Neurite radii increase linearly (upper traces), own spike sendings
(lower trace) result in instantaneous shrinkage. Spike arrivals increase
the instantaneous firing rate by the coupling strength, it decays
exponentially in between (middle trace). }
\end{figure}

\emph{Network growth.---}Neurons are commonly arranged in single
or stacked layers. We thus represent neurite extents by disks with
radii $R_{i}(t)$, with centers, representing cell somas, randomly
and uniformly distributed in a planar area \citep{OP94,abbott2007,barral2016synaptic}.
Since neurons with more neurite overlap can grow more synaptic connections
\citep{abeles1991corticonics,ooyen2014independently}, connection
strengths are set proportional to the overlap areas $A_{ij}(t)$ of
the disks, with proportionality constant $g$. We incorporate homeostatic
neurite growth by evolving extents as
\begin{equation}
\dot{R}_{i}(t)=K\left(1-\frac{1}{f_{\text{sat}}}\sum_{\hat{t}_{i}}\delta(t-\hat{t}_{i})\right),\label{hw_model_radius}
\end{equation}
Fig.~\ref{fig:MicroscopicModelDynamics}. Between spikes of neuron
$i$, $R_{i}(t)$ grows linearly with rate $K$. At spike sending,
it shrinks by a constant amount $K/f_{\text{sat}}$, which determines
the rate $f_{\text{sat}}$ at which growth and shrinkage equilibrate.
There are no self-connections. Growth takes much longer than decay
of activity, $1/K\gg\tau$ (spatial scales of the population are of
order one). Furthermore, we assume $f_{\text{sat}}\gg f_{0}$, in
agreement with experiments \citep{ford2012,yada2017development,Supplement18CritGrowth}\nocite{johansson1994single,ZLZ06}.
Spontaneously inactive neurons would reduce the relevant average $f_{0}$
\citep{Supplement18CritGrowth}. The growth model is biologically
inspired; it is a simplification of previous growth models \citep{OP94,abbott2007,TOEWB10,ooyen2017rewiring}.
However, many slow homeostatic processes \citep{Ooyen11,turrigiano2012homeostatic,Zierenberg2018}
with $f_{\text{sat}}\gg f_{0}$ will yield similar results. 

\begin{figure}
\includegraphics[width=1\columnwidth]{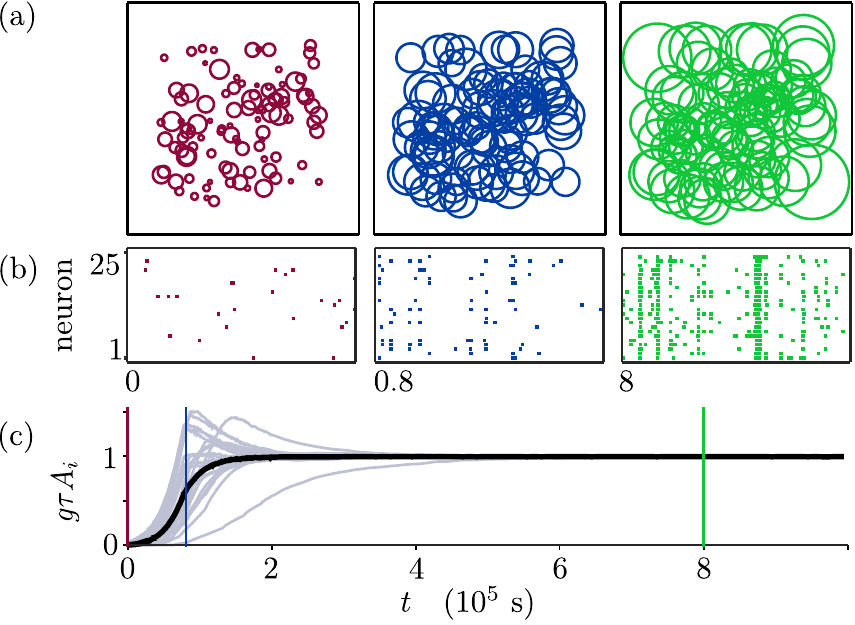} \caption{\label{fig:GlobalModelDynamics} Network dynamics. (a) Extents of
neurites. (b) Spikes generated by $25$ sample neurons ($100\,\text{s}$
windows). (c) Scaled total overlaps of 25 sample neurons (gray) and
the population average (black) as a function of time. For (a) and
(b) from left to right: initial state (red), state with growth on
average (blue), stationary state (green). Color coded vertical lines
in (c) indicate the three different time points in (a) and (b).}
\end{figure}

The neurons are initially mostly isolated. Over time, they extend
their neurites, form connections, and develop a network, Fig.~\ref{fig:GlobalModelDynamics}.
At intermediate stages, neurites and overlaps can overshoot \citep{OP94,TOEWB10,yada2017development}.
When neuron $i$'s time-averaged firing rate $\bar{f}_{i}$ reaches
$f_{\text{sat}}$, its average growth stops {[}Eq.~\eqref{hw_model_radius}{]}.
Our networks grow into a stationary state, where $\bar{f}_{i}=f_{\text{sat}}$
for all $i$. In the following, we investigate this state.

\emph{Stationary state dynamics.}---We first compute the average
number of spikes that a spike directly causes: Identical $\bar{f}_{i}$
imply identical time-averaged total overlaps $\sum_{j}\bar{A}_{ij}=\bar{A}_{i}=\bar{A}$
and input coupling strengths. Time averaging Eq.~\eqref{hw_model_rate},
$\bar{f}_{i}=f_{0}+\tau g\sum_{j}\bar{A}_{ij}\bar{f}_{j}$ {[}here
and henceforth we neglect the small fluctuations of $A_{ij}(t)$ around
$\bar{A}_{ij}${]}, and inserting $f_{\text{sat}}$ yields $\tau g\bar{A}=1-f_{0}/f_{\text{sat}}$.
A spike of neuron $j$ at $\hat{t}_{j}$ adds $gA_{ij}(\hat{t}_{j}^{-})e^{-(t-\hat{t}_{j})/\tau}\Theta(t-\hat{t}_{j})$
to $f_{i}(t)$ {[}Eq.~\eqref{hw_model_rate}, $\Theta$ is the Heaviside
function{]}, such that the number of additionally induced spikes in
neuron $i$ is Poisson distributed with mean $\tau gA_{ij}(\hat{t}_{j}^{-})$.
Averaged over the randomness of spike generation each spike thus generates
in total
\begin{align}
\sigma & =\tau g\sum_{i}\bar{A}_{ij}=\tau g\bar{A}=1-\frac{f_{0}}{f_{\text{sat}}}\label{eq:sigma}
\end{align}
spikes, where we used the symmetry of overlaps, $A_{ij}=A_{ji},\,\sum_{i}\bar{A}_{ij}=\sum_{j}\bar{A}_{ij}=\bar{A}$.
Equation \eqref{eq:sigma} holds independently of network activity
and neuron identity, due to the linearity of Eq.~\eqref{hw_model_rate}
and the homogeneity of parameters. In particular, $\sigma$ equals
also the time and population average number of induced spikes ($\propto f_{\text{sat}}-f_{0}$)
per spike ($\propto f_{\text{sat}}$).

The independence of spike offspring generation from other spikes allows
us to understand the dynamics as a branching process with branching
parameter $\sigma$. More specifically, we have an age-dependent or
Crump-Mode-Jagers branching process \citep{crump1968}: Individuals
(spikes) generate offspring at an age-dependent rate. Neuronal avalanches
are trees of offspring, started by a spontaneous spike. For their
overall size only the distribution of single spike offspring matters.
It is Poissonian with parameter $\sigma$. The avalanche sizes $s$
therefore follow the Borel distribution~\citep{tanner1961},
\begin{equation}
\begin{split}P(s) & =\frac{(s\sigma)^{s-1}e^{-s\sigma}}{s!}.\end{split}
\label{eq:Borel}
\end{equation}
We apply Stirling's approximation to obtain
\begin{equation}
P_{\text{appr}}(s)=\frac{1}{\sqrt{2\pi}\sigma}s^{-3/2}e^{-(\sigma-\ln\sigma-1)s},\label{eq:BorelStirling}
\end{equation}
explicitly highlighting the power-law tail with exponent $3/2$ of
a critical branching process for $\sigma=1$ \citep{ZLS95,Jensen98,harris2002theory,santo2017simple}.
For a subcritical process ($\sigma<1$), Eq.~\eqref{eq:BorelStirling}
is a power law with exponential cutoff around $s_{c}(\sigma)=(\sigma-\ln\sigma-1)^{-1}$.
It signifies subcritical dynamics \citep{Jensen98,ZLS95,priesemann2014spike},
not a finite size effect \citep{BTW88,Jensen98}; the size distribution
is independent of neuron number. Equation~\eqref{eq:BorelStirling}
inherits the good quality of Stirling's approximation \citep{abramowitz1972handbook},
with relative error about $1/(12s)$.

\begin{figure}
\noindent \begin{centering}
\includegraphics[width=1\columnwidth]{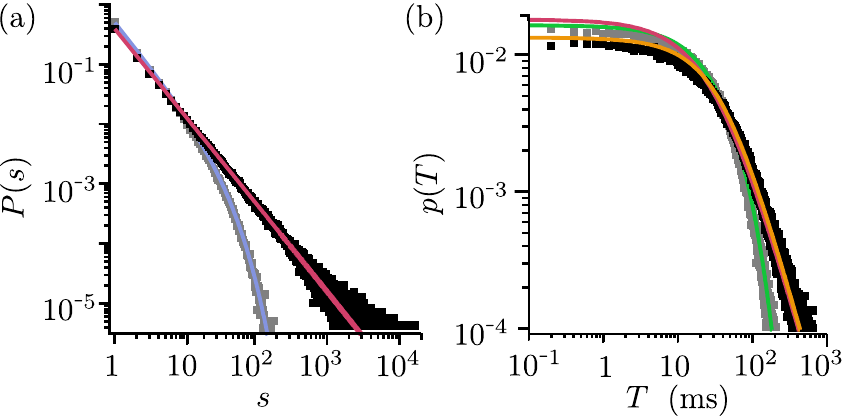}
\par\end{centering}
\caption{\label{fig:AvalancheSizesDurations} Avalanche sizes and durations.
(a) Analytical size distributions Eq.~\eqref{eq:BorelStirling} (discrete
points connected) and simulation results for subcritical ($f_{\text{sat}}=0.04\,\text{Hz},\,\sigma=0.75,\,t_{\text{bin}}=30\,\text{ms}$,
blue and gray) and near-critical ($f_{\text{sat}}=2\,\text{Hz},\,\sigma=0.995,\,t_{\text{bin}}=45\,\text{ms}$,
red and black) states. Equation \eqref{eq:Borel} yields visually
indistinguishable analytics. (b) Analytical duration distributions
Eq.~\eqref{eq:dur_p_exact} and simulation results, for subcritical
(green and gray) and near-critical (orange and black) states, and
closed-form approximation Eq.~\eqref{dur_p_approx} (red).}
\end{figure}

The heights of Crump-Mode-Jagers trees, i.e., the temporal differences
$T$ between their first and last individuals, represent the durations
of the corresponding neuronal avalanches. In the following we derive
their probability density $p(T)$. Because of the additivity of Poisson
processes, the superposition of all neurons' spike trains can be described
as an inhomogeneous Poisson process with rate $f(t)=\sum_{i}f_{i}(t)$.
Summing Eq.~\eqref{hw_model_rate} over $i$ and inserting $\bar{A}=\sigma/(g\tau)$
{[}Eq.~\eqref{eq:sigma}{]} yields $\tau\dot{f}(t)=Nf_{0}-f(t)+\sigma\sum_{\hat{t}}\delta(t-\hat{t})$
with the number of neurons $N$. $\hat{t}$ are the neurons' spike
times; they occur with instantaneous rate $f(t)$. The spiking dynamics
may thus be interpreted as a self-exciting Hawkes process. It is Markovian
due to the exponentially decaying impact kernel \citep{oakes1975markovian,bacry2015hawkes}.
The spontaneous background rate $Nf_{0}$ initiates avalanches. To
determine their durations, we consider an analogous process with instantaneous
rate $f_{a}(t)$ and without spontaneous spiking, which is initiated
at $t=0$ by an external spike,
\begin{equation}
\tau\dot{f}_{a}(t)=-f_{a}(t)+\sigma\sum_{\hat{t}_{a}}\delta(t-\hat{t}_{a}),\,\,f_{a}(0)=\frac{\sigma}{\tau}.\label{eq:ftnobackground-1}
\end{equation}
The duration of an avalanche is the time $T$ of this process' last
spike. The probability that it has occurred before $t$ gives the
distribution function $P(T\leq t)$ of durations. We first compute
this probability conditioned on the instantaneous rate $f_{a}(t)$
at the end of the considered interval:
\begin{equation}
\begin{split}P(T\leq t\mid f_{a}(t)) & =P(\textrm{no spike in }(t,\infty)\mid f_{a}(t))\\
 & =e^{-\int_{t}^{\infty}f_{a}(t')dt'}=e^{-\tau f_{a}(t)},
\end{split}
\label{eq:PTCond}
\end{equation}
where we use that the process behaves like a Poisson process with
exponentially decaying rate, if no spike is generated. Averaging over
$f_{a}(t)$ yields
\begin{equation}
\begin{split}P(T\leq t) & =\int_{0}^{\infty}P\left(T\leq t\mid f_{a}(t)\right)p\left(f_{a}(t)\right)df_{a}(t)\\
 & =E\bigl(e^{-\tau f_{a}(t)}\bigr).
\end{split}
\label{eq:PTUncond}
\end{equation}
$E(\cdot)$ denotes the expectation value over the process, Eq.~\eqref{eq:ftnobackground-1}.
Importantly, Eq.~\eqref{eq:PTUncond} shows that $P(T\leq t)$ equals
the Laplace transform of the random variable $f_{a}(t)$, evaluated
at the decay time $\tau$. This Laplace transform has recently been
derived \citep{gau2015,dassios2011dynamic,errais2010affine}. Inserting
our parameters yields $E(e^{-\tau f_{a}(t)})=e^{\sigma a(t)/\tau}$,
where $a(t)$ satisfies 
\begin{equation}
\dot{a}(t)=-a(t)/\tau+e^{\sigma a(t)/\tau}-1,\,\,a(0)=-\tau.\label{eq:a_ode}
\end{equation}
The resulting $P(T\leq t)=e^{\sigma a(t)/\tau}\Theta(t)$ with $\Theta(0)=1$
has the density
\begin{equation}
p(T)=\sigma\dot{a}(T)e^{\sigma a(T)/\tau}\Theta(T)/\tau+e^{-\sigma}\delta(T).\label{eq:dur_p_exact}
\end{equation}
We can generalize Eq.~\eqref{eq:a_ode} to Hawkes processes with
different kernels using the integral equation for cluster duration
distributions \citep{hawkes1974cluster,moller2005}.

We finally approximate $p(T)$ by closed-form expressions with a focus
on its tail near criticality. For large $t$, $P(T\leq t)$ approaches
$1,$ so $a(t)$ approaches $0$. Generally, $\sigma a(t)/\tau$ stays
between $-1$ and $0$. Expanding $e^{\sigma a(t)/\tau}$ in Eq.~\eqref{eq:a_ode}
around $\sigma a(t)/\tau=0$ to second order, 
\begin{equation}
\dot{a}(t)\approx(\sigma-1)a(t)/\tau+\sigma^{2}a(t)^{2}/(2\tau^{2}),\,\,a(0)=-\tau,\label{eq:adotapprox}
\end{equation}
yields closed-form approximations for $a(t)$. In particular, for
nearly critical systems with $\sigma\approx1$, the first term on
the right-hand side vanishes and the solution is $a_{\text{appr}}(t)=-2\tau^{2}/(2\tau+t)$,
leading to a probability density 
\begin{equation}
p_{\text{appr}}(T)=2\tau(2\tau+T)^{-2}e^{a_{\text{appr}}(T)/\tau}\Theta(T)+e^{-1}\delta(T),\label{dur_p_approx}
\end{equation}
which approaches for large $T$ a power law with critical exponent
$2$. For large $t$ the error in the expansion Eq.~\eqref{eq:adotapprox}
becomes negligible, $a_{\text{appr}}(t)$ thus has the right slope
and $p_{\text{appr}}(T)$ equals $p(T)$ up to a factor (Fig.~\ref{fig:AvalancheSizesDurations}).
We conclude that the duration distribution has a power-law tail with
critical exponent $2$. Expanding the exponential to third order yields
a closed-form distribution that is a good approximation also for small
$T$.

\emph{Simulations.---}We complement our analytics with simulations
to (i) compare the avalanche distributions, (ii) exemplify the irrelevance
of connectivity fluctuations, (iii) investigate the spatial spread
of avalanches, (iv) address the robustness of the results, and (v)
consider a typical experimental manipulation. If not stated otherwise,
$N=100$, $\tau=10\,\text{ms}$ \citep{DA01}, $g=500\,\text{Hz},\,f_{0}=0.01\,\text{Hz},\,f_{\text{sat}}=2\,\text{Hz}$
\citep{ford2012,yada2017development}, somas are placed on unit square,
$K^{-1}=10^{6}\,\text{s}$ (quick growth, accelerating simulations)
\citep{Ooyen11,turrigiano2012homeostatic,TOEWB10,yada2017development}.
The simulations use an event-based algorithm. Next spike times are
determined using inverse transform sampling of the interspike-interval
distribution; we avoid nonelementary functions by splitting each neuron's
Poisson process into a homogeneous (rate $f_{0}$) and an inhomogeneous
one. 

An avalanche should be a sequence of offspring spikes of one spontaneous
progenitor. To keep contact with the experimental literature, we analyze
numerical data by binning time and considering spike sequences that
are not separated by an empty bin as one avalanche \citep{BP03,priesemann2009subsampling,hahn2010neuronal,levina2017subsampling}.
Our model yields analytical estimates for the probabilities that binning
splits the first avalanche spikes or merges them with the next avalanche,
as well as for splitting or merging an average avalanche. Keeping
them moderate provides our bin sizes $t_{\text{bin}}$ in terms of
experimentally accessible quantities ($f_{0},\tau,N,f_{\text{sat}}$)
\citep{Supplement18CritGrowth}. Results are robust against changing
$t_{\text{bin}}$. 

(i) In all simulations the model reaches a stationary state. The avalanche
distributions agree well with the analytically derived ones, Fig.~\ref{fig:AvalancheSizesDurations},
the effects of binning and avalanche overlaps are small. We quantitatively
test this agreement using the methods described in Refs. \citep{clauset2009,aksenov2003}.
For both size and duration distributions a pure power law is ruled
out, as expected. For the size distribution, a power law with exponential
cutoff, cf.~Eq.~\eqref{eq:BorelStirling}, yields a good fit. The
analytical values of the power-law exponent, the cutoff $s_{c}(\sigma)$,
and the resulting $\sigma$ are closely matched.

(ii) The fluctuations of $\sum_{j}A_{ij}(t)$ and the deviations of
$\bar{f}_{i}$ from $f_{\text{sat}}$ are small ($<1\%$, Fig.~\ref{fig:GlobalModelDynamics}).
Freezing the network ($K=0$) in the stationary state has no effect
on the avalanche statistics: Neuronal growth carries the system close
to a critical point, but is not required later on. This is in agreement
with self-organized criticality and excludes other mechanisms \citep{sornette1994,bonachela2009self,bonachela2010self}.

\begin{figure}
\includegraphics[width=1\columnwidth]{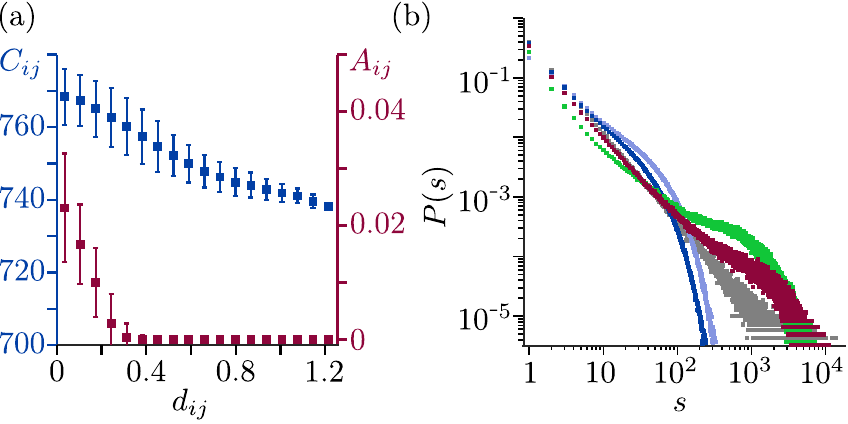} \caption{\label{fig:CorrelationLengthRefractoryPeriod} (a) Pairwise spike
number covariances and overlaps as functions of the intersomatic distances
$d_{ij}$ (averages around a particular intersomatic distance, bars:
standard deviations), $\sigma=0.995,\,t_{\text{bin}}=45\,\text{ms}$.
(b) Avalanche size distributions for the same model as in Fig.~3,
but with absolute refractory period $\tau_{\text{ref}}=\tau$ (red),
$\tau_{\text{ref}}=4\tau$ (green), $\tau_{\text{ref}}=0\,\text{ms}$
(gray) for reference, $t_{\text{bin}}=45\,\text{ms}$; $\tau=5\,\text{ms},\,\tau_{\text{ref}}=4\tau,\,f_{0}=0.1\,\text{Hz},\,f_{\text{sat}}=0.8\,\text{Hz},\,t_{\text{bin}}=10\,\text{ms}$
(blue, $t_{\text{bin}}=45\,\text{ms}$: light blue).}
\end{figure}

(iii) To investigate spatial spread near criticality, we compute covariances
$C_{ij}=\left\langle n_{i}n_{j}\right\rangle -\left\langle n_{i}\right\rangle \left\langle n_{j}\right\rangle $
between numbers $n_{i},\,n_{j}$ of spikes contributed to single avalanches
by neurons $i,\,j$ with various distances. Covariances decay comparably
slowly. Covariances and thus avalanches spread beyond direct connections,
Fig.~\ref{fig:CorrelationLengthRefractoryPeriod}(a).

(iv) To test robustness, we first freeze the growth in the stationary
state and shuffle the output vectors (columns of the coupling matrix)
between neurons. While this alters the network topology and breaks
coupling symmetry, it leaves the essential total coupling strengths
unchanged. Indeed, we observe little effect on avalanche sizes and
durations. Second, we consider moderate nonadditivity of spike impacts.
We introduce an absolute refractory period $\tau_{\text{ref}}$ after
a sent spike, during which the neuron cannot spike again. We observe
that although the refractory period limits the firing rate, the network
reaches a stationary state with the same average individual rate $f_{\text{sat}}$
as before: larger overlaps compensate refractoriness. For a refractory
period about $\tau$, which is often biologically plausible \citep{koch1999biophysics,DA01},
the statistics resemble the original one for small and medium size
avalanches {[}Fig.~\ref{fig:CorrelationLengthRefractoryPeriod}(b),
red vs.~gray{]}. Larger couplings and stop of avalanches lacking
available neurons cause an excess of larger avalanches, followed by
a strong reduction. Neurons still frequently contribute several spikes
to avalanches. With long refractoriness, little similarity remains
{[}Fig.~\ref{fig:CorrelationLengthRefractoryPeriod}(b), green and
blue; blue: our model with parameters adapted from Ref. \citep{abbott2007},
calcium variable present in Ref. \citep{abbott2007} does not affect
distribution shape{]}.

(v) Manipulation of neural excitability or coupling strength via $g$
causes subcriticality (decreased $g$) or excess of large avalanches
(increased $g$), as in experiments \citep{BP03,shew2011information}.
Our model predicts that the latter is balanced by network plasticity
faster than the former, due to strongly increased activity \citep{Supplement18CritGrowth}.

\emph{Discussion and conclusion.---}We suggest an analytically tractable
model for neural network growth, which may explain the emergence of
subcritical and critical avalanche dynamics. It covers essential features
of biological neurons such as operation in continuous time, spiking,
leak currents, and network growth. Still, it allows the analytical
computation of the avalanche size and duration distributions for subcritical
and critical stationary states. Our numerical analysis confirms their
validity and robustness and yields additional insight.

Two features are responsible for the emergence of the (near-)critical
state (Fig.~\ref{fig:AvalancheSizesDurations}): (i) homeostatic
growth to attain a saturation rate that is high compared to the spontaneous
one (precise values of $f_{\text{sat}}$ and $f_{0}$ are irrelevant),
and (ii) linear summation of spike impacts. (i) implies that in the
stationary state on average each spike generates nearly one successor.
This holds for all networks with largely self-sustained activity.
Usually, however, branching parameters vary, for example at high network
activity spikes generate less offspring. This drives activity excursions
back and generates non-power-law distributions \citep{VA05,KSAR07}.
In our networks, (ii) implies that the branching parameter is the
same for each spike. Small saturation rates yield subcritical dynamics
(Fig.~\ref{fig:AvalancheSizesDurations}), strong nonlinearities
other deviations {[}Fig.~\ref{fig:CorrelationLengthRefractoryPeriod}(b){]}.
Our model thus predicts that neural networks may develop criticality
already due to their growth, that spontaneous spiking in such networks
is low compared to saturation, and that spike effects add rather linearly
and are independent of activity level. For example, starburst amacrine
cells have radial dendritic trees, interact during development via
dendro-dendritic excitatory connections and are reported to generate
critical avalanches \citep{hennig2009}. Our model predicts that higher
precision measurements will reveal deviations as in Fig.~\ref{fig:CorrelationLengthRefractoryPeriod}(b),
due to the cells' long refractory periods. 

Our network model is based on the neurobiologically more detailed
ones \citep{OP94,abbott2007,TOEWB10}. Motivated by experiments, Ref.
\citep{OP94} proposes radial activity dependent neurite outgrowth
steered by calcium dynamics and finds convergence to a stationary
state for certain parameter ranges. To study avalanches, Ref. \citep{abbott2007}
adds stochastically spiking neurons, albeit with long refractoriness
and larger $f_{0}/f_{\text{sat}}$, impeding analytical treatment
and causing large deviations from criticality {[}Fig.~\ref{fig:CorrelationLengthRefractoryPeriod}(b){]};
Refs. \citep{TOEWB10} assumes antagonistic growth of axons and dendrites
and finds criticality, if a certain fraction of neurons becomes inhibitory;
Refs. \citep{levina2007criticality,droste2013analytical,Zierenberg2018}
consider more abstract homeostasis and neuron models.

Usually, models for neuronal avalanches only allow to estimate size
and duration distributions numerically \citep{LHG07,abbott2007,hennig2009,TOEWB10,delpapa2017}.
Reference \citep{EHE02} obtains an analytical expression of the size
distribution for a discrete-time network. Our article derives size
and duration distributions for a continuous-time spiking network model
after self-organization. These distributions depend only on the experimentally
accessible parameters $f_{0}/f_{\text{sat}}$ and $\tau$ (duration
scaling). The power-law exponents agree with experimentally found
ones and those of simple branching processes \citep{BP03,hennig2009,ZLS95,harris2002theory}.
The duration distribution has power-law scaling at the tail \citep{hennig2009,ZLS95,harris2002theory},
a fit to short avalanches \citep{BP03} would yield different results.
Our analytical expressions allow fast parameter scans, delineations
of the (near-)critical regime and parameter estimations.

From a general perspective, avalanches in our model are clusters of
a Hawkes process. While their size distribution can be straightforwardly
computed {[}Eq.~\eqref{eq:Borel}{]}, their duration distribution
generally requires solving a nonlinear integral equation \citep{hawkes1974cluster,moller2005}.
Here we show that for Markovian Hawkes processes it follows from the
solution of an ordinary differential equation {[}Eqs.~\eqref{eq:a_ode},\eqref{eq:dur_p_exact}{]}
and we give closed-form approximations. This may find straightforward
application in further fields of science where these processes are
employed, for example, to characterize durations of financial market
fluctuations \citep{errais2010affine}, earthquakes \citep{wang2012markov},
violence \citep{lewis2012}, and epidemics \citep{kim2011,Schoenberg2017}.
\begin{acknowledgments}
We thank Matthias Hennig, Anna Levina, Viola Priesemann, and Johannes
Zierenberg for helpful discussions and the German Federal Ministry
of Education and Research (BMBF) for support via the Bernstein Network
(Bernstein Award 2014, 01GQ1501 and 01GQ1710).
\end{acknowledgments}

\providecommand{\noopsort}[1]{}\providecommand{\singleletter}[1]{#1}%

\end{document}


\title{Growing Critical: Self-Organized Criticality in a Developing Neural
System\\
\textendash{} Supplemental Material \textendash{}}

\author{Felipe Yaroslav Kalle Kossio}

\author{Sven Goedeke}

\affiliation{Neural Network Dynamics and Computation, Institute of Genetics, University
of Bonn, Bonn, Germany}

\author{Benjamin van den Akker}

\affiliation{Department of Neuroinformatics, Radboud University Nijmegen, Nijmegen,
Netherlands}

\author{Borja Ibarz}

\affiliation{Nonlinear Dynamics and Chaos Group, Departamento de Fisica, Universidad
Rey Juan Carlos, Madrid, Spain}

\author{Raoul-Martin Memmesheimer}

\affiliation{Neural Network Dynamics and Computation, Institute of Genetics, University
of Bonn, Bonn, Germany}

\affiliation{Department of Neuroinformatics, Radboud University Nijmegen, Nijmegen,
Netherlands}
\maketitle

\section{Manipulation of neural excitability}

\begin{figure}
\begin{centering}
\includegraphics[width=0.75\columnwidth]{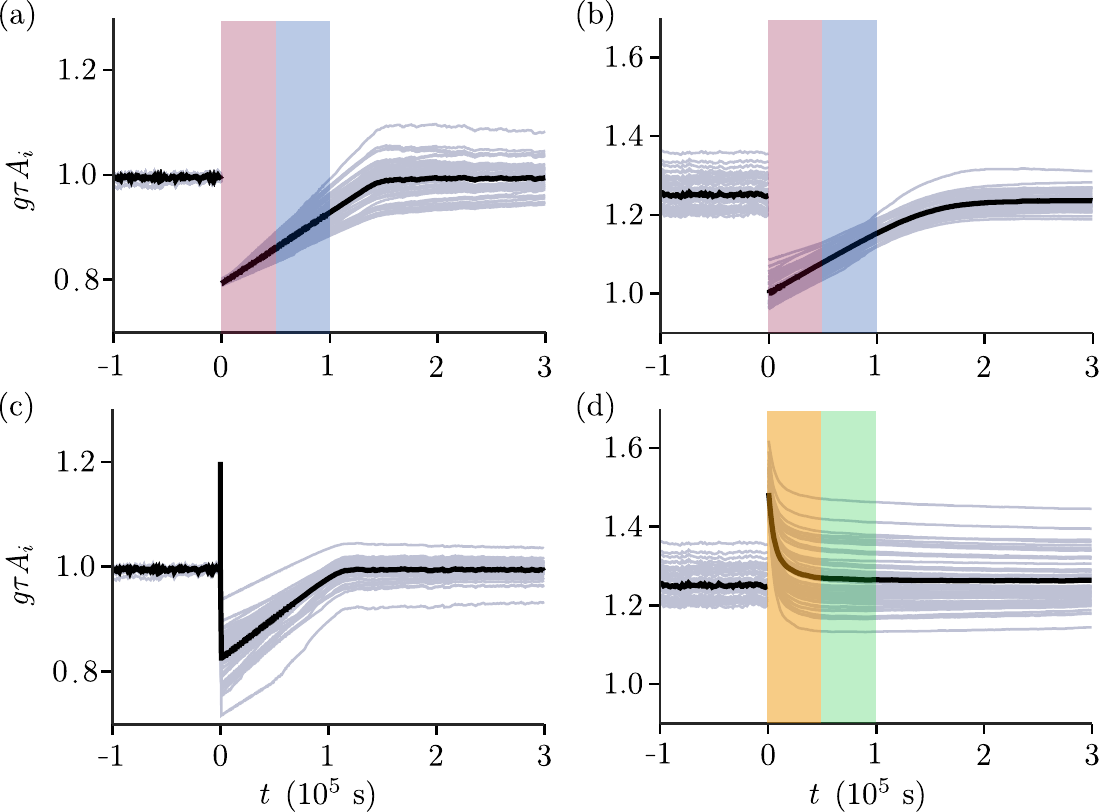}
\par\end{centering}
\caption{Coupling strengths after manipulation of neural excitability. Scaled
total overlaps of 50 neurons (gray) and mean total overlap (black),
similar to Fig.~2(c). At $t=0\,\text{s}$, the excitability $g$
of all neurons is decreased, $g\rightarrow0.8g$ (a,b), or increased,
$g\rightarrow1.2g$ (c,d). Before, the networks are in a stationary
state. Neurons in (b,d) have refractory period $\tau_{\text{ref}}=\tau$.
Avalanche size distributions for the color shaded areas are displayed
in Fig.~\ref{fig:FigB} (sampling time $0.5\times10^{5}\,\text{s}$
each). The network growth rate is set to $K^{-1}=2\times10^{7}\,\text{s}$
prior to manipulation to allow longer sampling times.\label{fig:FigA}}
\end{figure}

\begin{figure}
\begin{centering}
\includegraphics[width=1\columnwidth]{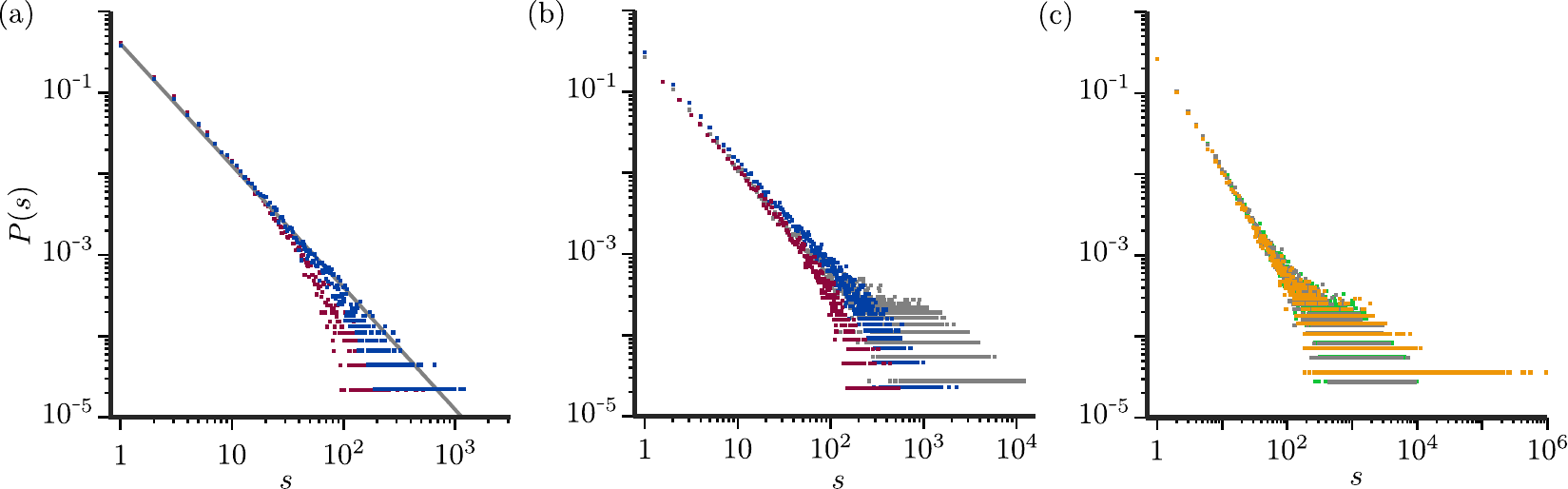}
\par\end{centering}
\caption{Avalanche size distributions after manipulation of neural excitability.
(a) Network of Fig.~\ref{fig:FigA}(a). Red and blue: distributions
sampled directly after decreasing excitability {[}Fig.~\ref{fig:FigA}(a),
red shading{]} and in the subsequent interval {[}Fig.~\ref{fig:FigA}(a),
blue shading{]}, respectively. Gray: analytical stationary state distribution
Eq.~(4). (b) Like (a) for the network of Fig.~\ref{fig:FigA}(b).
Gray: stationary state distribution (sampled from simulation). (c)
Network of Fig.~\ref{fig:FigA}(d). Orange: distribution sampled
directly after increasing excitability {[}Fig.~\ref{fig:FigA}(d),
orange shading{]}. Green: distribution sampled during the subsequent
interval {[}Fig.~\ref{fig:FigA}(d), green shading{]}. Gray: stationary
state distribution.\label{fig:FigB}}
\end{figure}

To address the impact of a typical experimental manipulation on the
dynamics of our networks, we change the neural excitability or, equivalently,
the coupling strength via $g$. We find that after decreasing it,
in the short term activity is subcritical, Figs.~\ref{fig:FigA}(a),
\ref{fig:FigB}(a). The neurites react to the resulting overall loss
of input by outgrowth, which leads to strengthening of connections
and finally to the recovery of the near-critical state. An increase
of neural excitability leads to very strong activity, which is quickly
overcompensated by a decrease of connection strengths within the course
of a single large avalanche, Fig.~\ref{fig:FigA}(c). The system
becomes subcritical and regains the near-critical state more slowly
thereafter. In biological neural networks overly large spiking activity
is prevented by refractoriness. When decreasing neural excitability,
our network models incorporating refractoriness behave similar to
those without, Figs.~\ref{fig:FigA}(b), \ref{fig:FigB}(b). When
increasing excitability, in the short term they display an excess
of medium size and large avalanches, Fig.~\ref{fig:FigB}(c). The
overlap sizes and coupling strengths decrease until the near-critical
state is regained, Fig.~\ref{fig:FigA}(d). This happens faster than
the adaptation from subcriticality due to the still large excess of
spiking: in Fig.~\ref{fig:FigB}(c) the distribution in green is
already similar to that in gray, in Fig.~\ref{fig:FigB}(a,b) the
distributions in blue are markedly different from those in gray. 

In agreement with our findings, experimental studies have shown that
a global decrease of excitatory synaptic strengths (decrease of network
excitability) leads to subcritical activity while a global decrease
of inhibitory synaptic strengths (increase of network excitability)
leads to supercritical behavior with an excess of large avalanches
\citep{BP03,shew2011information}.

\section{Robustness of avalanche characteristics against changes in the spontaneous
and saturation spike rates}

\begin{figure}
\begin{centering}
\includegraphics[width=0.7175\columnwidth]{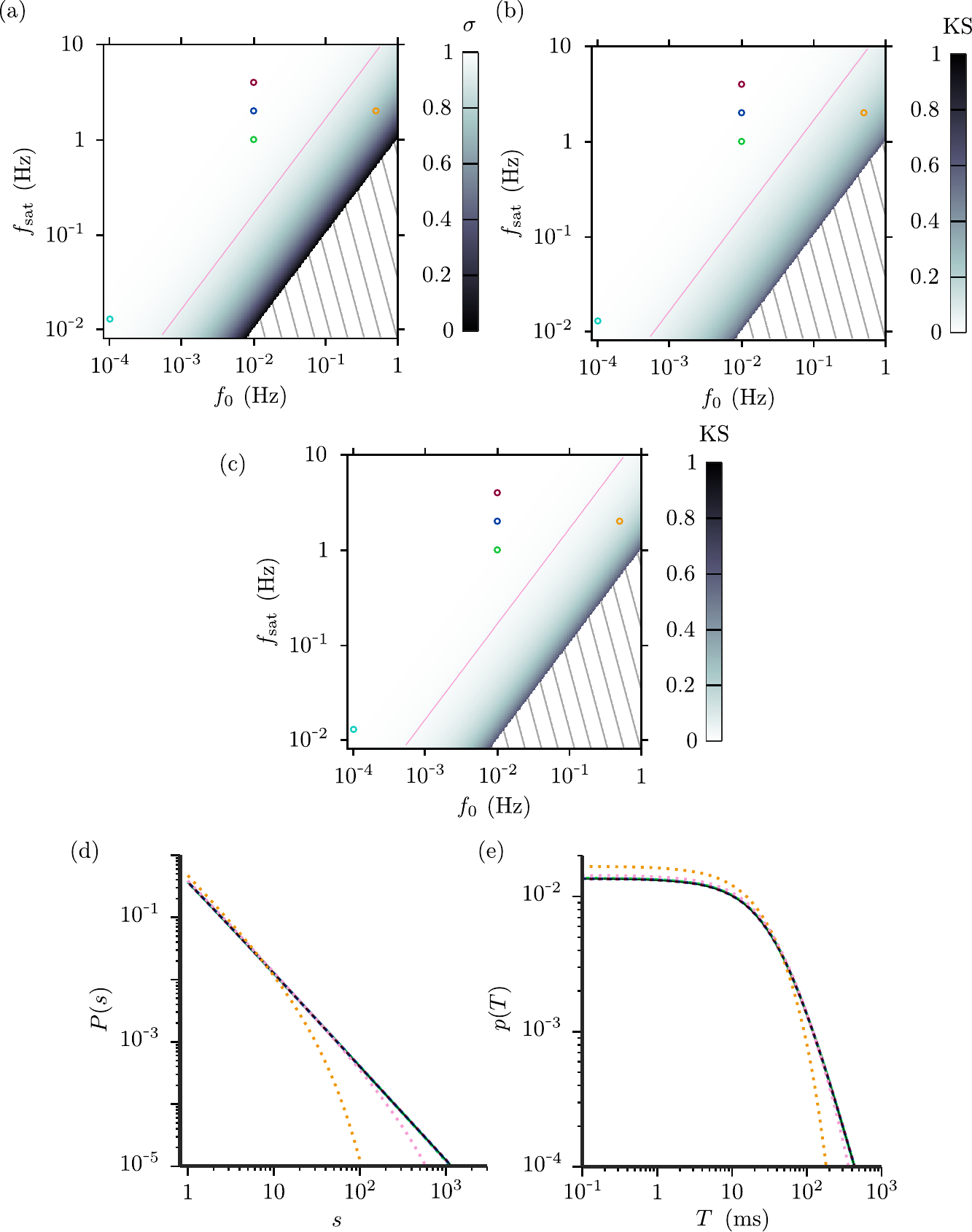}
\par\end{centering}
\caption{Stationary state avalanche size and duration distributions are near-critical
for large ranges of $f_{\text{sat}}$ and $f_{0}$. (a) Branching
parameter $\sigma$ (gray scale, hatched area: $\sigma<0$, system
undefined) of the stationary state dynamics as a function of $f_{0}$
and $f_{\text{sat}}$. $\sigma$ is close to $1$ for large ranges
of $f_{\text{sat}}$ and $f_{0}$. (b) Kolmogorov-Smirnov distance
(KS) between the critical size distribution {[}Eq.~(4), $\sigma=1${]}
and distributions with different $f_{\text{sat}}$ and $f_{0}$ {[}Eq.~(4),
$\sigma=1-f_{0}/f_{\text{sat}}${]}. The distance is close to $0$
for large ranges of $f_{\text{sat}}$ and $f_{0}$. (c) Like (b) for
the avalanche duration distributions Eqs.~(9), (10). KS is close
to $0$ for large ranges of $f_{\text{sat}}$ and $f_{0}$. Circles:
$f_{\text{sat}}=2\,\text{Hz},\,f_{0}=0.01\,\text{Hz}$ (blue, $\sigma=0.995$,
used in the main text), $f_{\text{sat}}=4\,\text{Hz},\,f_{0}=0.01\,\text{Hz}$
(red, $\sigma=0.9975$), $f_{\text{sat}}=1\,\text{Hz},\,f_{0}=0.01\,\text{Hz}$
(green, $\sigma=0.99$), $f_{\text{sat}}=13\,\text{mHz},\,f_{0}=0.1\,\text{mHz}$
(cyan, $\sigma=0.9923$) \citep{ford2012}, $f_{\text{sat}}=2\,\text{Hz},\,f_{0}=0.5\,\text{Hz}$
(orange, $\sigma=0.75$); pink line: $f_{0}/f_{\text{sat}}=0.06$,
upper limit of the ratio derived from Ref. \citep{yada2017development}
($\sigma=0.94$). (d,e) Size and duration distributions for the parameter
values highlighted in (a-c) in alike colors (dotted for better discrimination,
$\tau=10\,\text{ms}$). The distributions in red, blue, green, cyan,
and pink are near-critical, they partially overlay each other and
the critical distributions (black). The distributions in orange are
subcritical.\label{fig:FigE}}
\end{figure}

The stationary state avalanche size and duration distributions are
largely independent of the choice of $f_{0}$ and $f_{\text{sat}}$,
Fig.~\ref{fig:FigE}. They are practically critical whenever $f_{0}$
is small against $f_{\text{sat}}$. For all finite values of $f_{\text{sat}}$
and non-zero values of $f_{0}$ there is ultimately an exponential
cutoff, see Eq.~(5) and Eqs.~(9), (10) after expanding around $a(t)=0$.
The large range parameter scans in Fig.~\ref{fig:FigE} are greatly
facilitated by our analytical formulas: They allow us to efficiently
determine the avalanche characteristics for the markedly subcritical
as well as for the near-critical regime, where usually long numerical
simulations are necessary to capture the distribution tails with their
large and long avalanches. For illustration, Fig.~\ref{fig:FigE}
exemplarily highlights the results for the parameters used in Fig.~3
as well as for parameter sets where $f_{\text{sat}}$ is multiplied
or divided by two (red, green, distributions are near-critical), or
$f_{0}$ is increased by a factor of 50 (orange, distribution is markedly
subcritical).

Values of $f_{0}$ and $f_{\text{sat}}$ were directly measured in
neural systems generating neuronal avalanches. The findings confirm
that our assumption of $f_{0}/f_{\text{sat}}\ll1$ is justified: Ref.
\citep{ford2012} measures spike rates in retinal starburst amacrine
cells in isolation and embedded in their network, where critical avalanches
were reported a few days after birth \citep{hennig2009}. The study
finds spontaneous spike rates of isolated neurons that decrease from
$f_{0}\approx1.3\,\text{mHz}$ to $f_{0}\approx0.1\,\text{mHz}$ between
postnatal day 2 and postnatal day 6, Fig.~1 in Ref. \citep{ford2012}.
The average spike rate $\bar{f}\approx13\,\text{mHz}$ of connected
neurons stays constant; we may assume $f_{\text{sat}}\approx13\,\text{mHz}$
(or higher, if the homeostatic network plasticity is slow). The ratio
$f_{0}/f_{\text{sat}}$ therefore decreases from $f_{0}/f_{\text{sat}}\approx0.1$
to $f_{0}/f_{\text{sat}}\approx0.01$, Fig.~\ref{fig:FigE}, cyan.
Reference \citep{yada2017development} investigates avalanches in
neuronal cultures. The study reports a population spike rate, which
rises from approximately $45\,\mathrm{Hz}$ at day 4 in vitro to approximately\textcolor{black}{{}
$730\,\mathrm{Hz}$ a}t day 30 in vitro, Fig.~5 in Ref. \citep{yada2017development}.
The first value is an upper estimate for $f_{0}\times N$, where $N$
is the number of neurons recorded from. The estimate neglects already
existing couplings and mutual excitation between neurons after four
days. The second value is a lower estimate for $f_{\text{sat}}\times N$,
since the networks may grow further. We thus have $f_{0}<45\,\text{Hz}/N$
and $f_{\text{sat}}>730\,\text{Hz}/N$, such that $f_{0}/f_{\text{sat}}<0.06$,
Fig.~\ref{fig:FigE}, pink line.

\section{Independence of avalanche characteristics of other model parameters}

Equation~(4) shows that the stationary state avalanche size distribution
only depends on $f_{0}$ and $f_{\text{sat}}$ (via $\sigma=1-f_{0}/f_{\text{sat}}$).
Similarly, Eqs.~(9), (10) show that the duration distribution depends
only on $f_{0}$, $f_{\text{sat}}$, and $\tau$. Equation~(6) implies
that changing $\tau$ is equivalent to rescaling the time axis and
therefore only leads to a rescaling of the duration distribution.
The dependence of avalanche statistics on the other model parameters,
which do not enter Eqs.~(4), (9), and (10), such as the coupling
parameter $g$, the number of neurons $N$, the growth rate $K$,
the dimensions of the square where the neurons are placed, and the
way they are placed in (regular vs.~random) vanishes during the network's
self-organization process. 

\begin{figure}
\begin{centering}
\includegraphics[width=0.75\columnwidth]{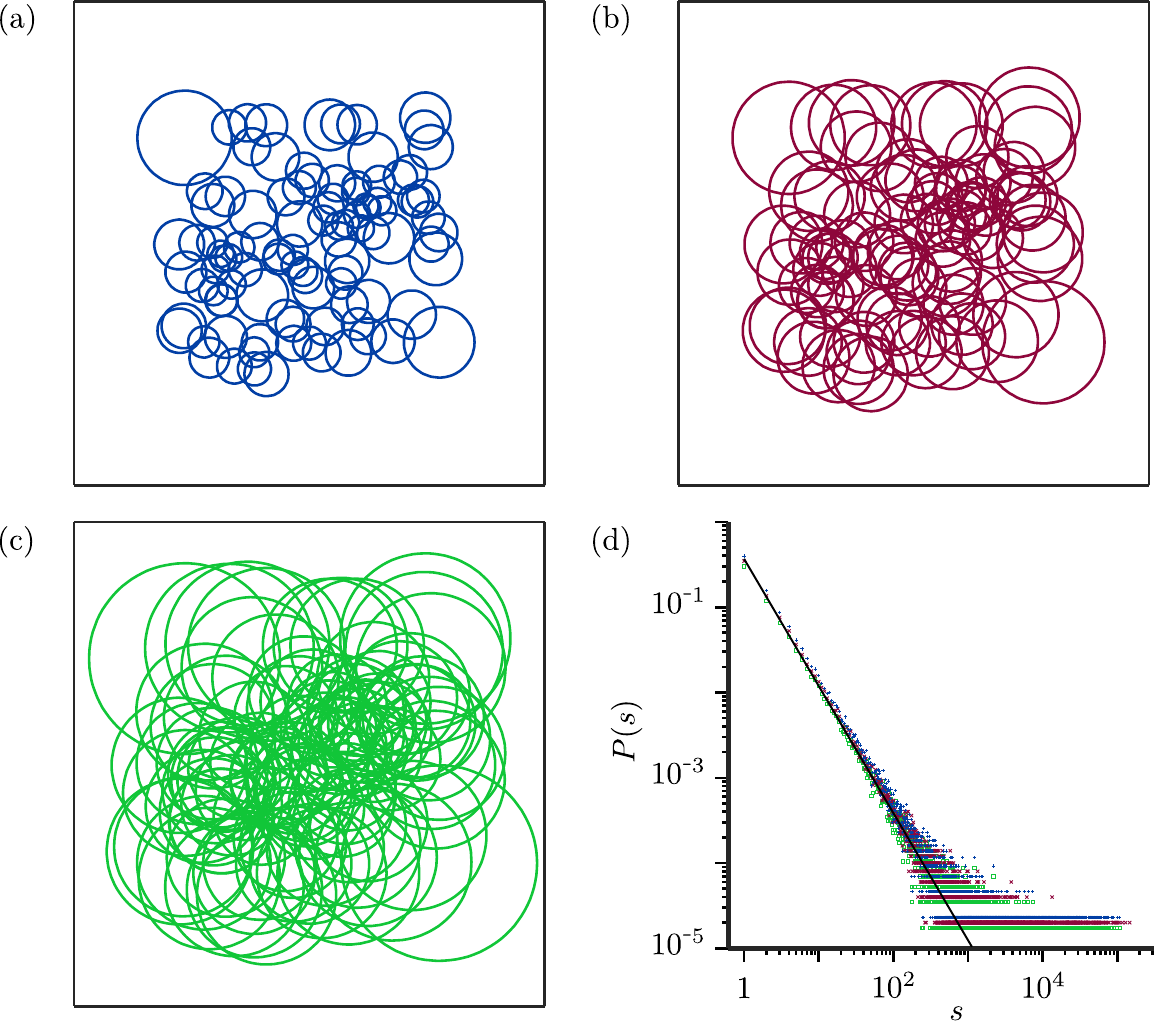}
\par\end{centering}
\caption{The choice of the parameter $g$ does not influence the avalanche
statistics in the stationary state. (a), (b), and (c) display networks
of $N=100$ neurons in the stationary state for $g=5\,\text{kHz}$,
$g=500\,\text{Hz}$, and $g=50\,\text{Hz}$, respectively. (d) shows
that the avalanche size distributions of these networks agree {[}networks
in (a), (b), (c): blue +, red x, green squares{]}. Distributions are
slightly vertically shifted for better discriminability.\label{fig:FigG}}
\end{figure}
 Consider as an example the coupling parameter $g$: If $g$ is small,
the network growth leads to larger total overlaps involving more synaptic
partners. If $g$ is large, the network growth leads to smaller total
overlaps involving fewer synaptic partners. In the end, the dynamics
are near-critical irrespective of $g$'s value, Fig.~\ref{fig:FigG}.
The value of $g$ may thus be chosen according to the biological system
to which our model is applied. Indeed, ranges of overlaps and numbers
of synaptic connection partners differ widely in full grown networks
of potential relevance for our model: Neurons in cultures establish
connections to tens and hundreds of other neurons, depending on the
density of the plating \citep{barral2016synaptic}, starburst amacrine
cells receive inputs from tens of other starburst amacrine cells \citep{ZLZ06,hennig2009},
and in the intact cortex, neurons have thousands of different neurons
as synaptic partners \citep{abeles1991corticonics}.

\section{Spontaneously active subpopulation}

\begin{figure}
\begin{centering}
\includegraphics[width=0.75\columnwidth]{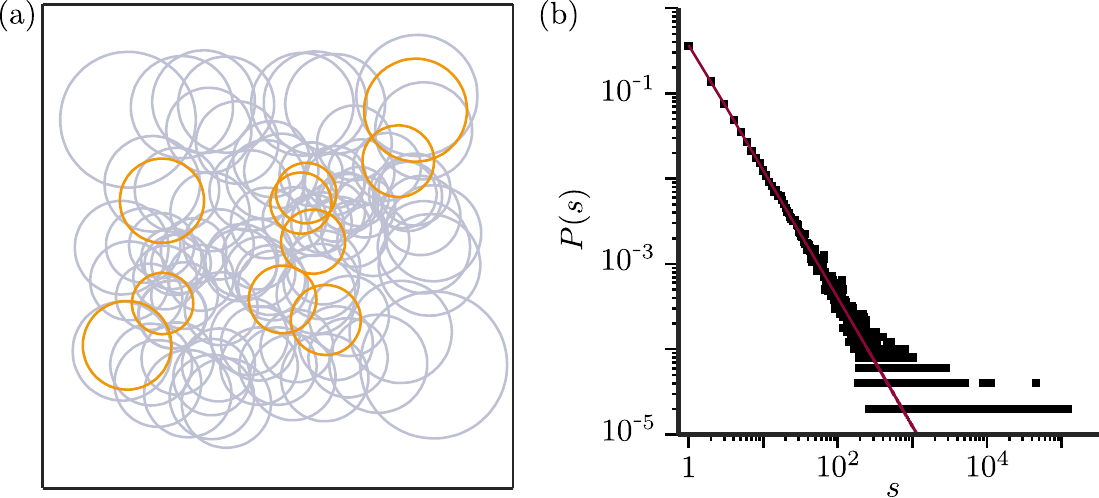}
\par\end{centering}
\caption{Avalanche statistics in the stationary state are unchanged, if only
a subpopulation of neurons is spontaneously active. (a) Network of
neurons in the stationary state with 10 out of 100 neurons spontaneously
active with rate $\tilde{f}_{0}=0.1\,\mathrm{Hz}$ (orange). (b) The
network's avalanche size distribution (numerically estimated, black)
agrees with that of a network where all neurons are spontaneously
active with $f_{0}=0.1\tilde{f}_{0}$ {[}Eq.~(4), red{]}.\label{fig:SubpopSpontA}}
\end{figure}

The stationary state avalanche size and duration distributions in
our model are unchanged, if only a subpopulation of neurons is spontaneously
active \citep{johansson1994single}. The quantity $f_{0}$ relevant
for the statistics is the average spontaneous activity per neuron,
see Fig.~\ref{fig:SubpopSpontA}. The size of the spontaneously active
subpopulation may be small against $N$, which leads to an $f_{0}$
that is small against the individual spontaneous rates.

\section{Binning}

\begin{figure}
\begin{centering}
\includegraphics[width=0.75\columnwidth]{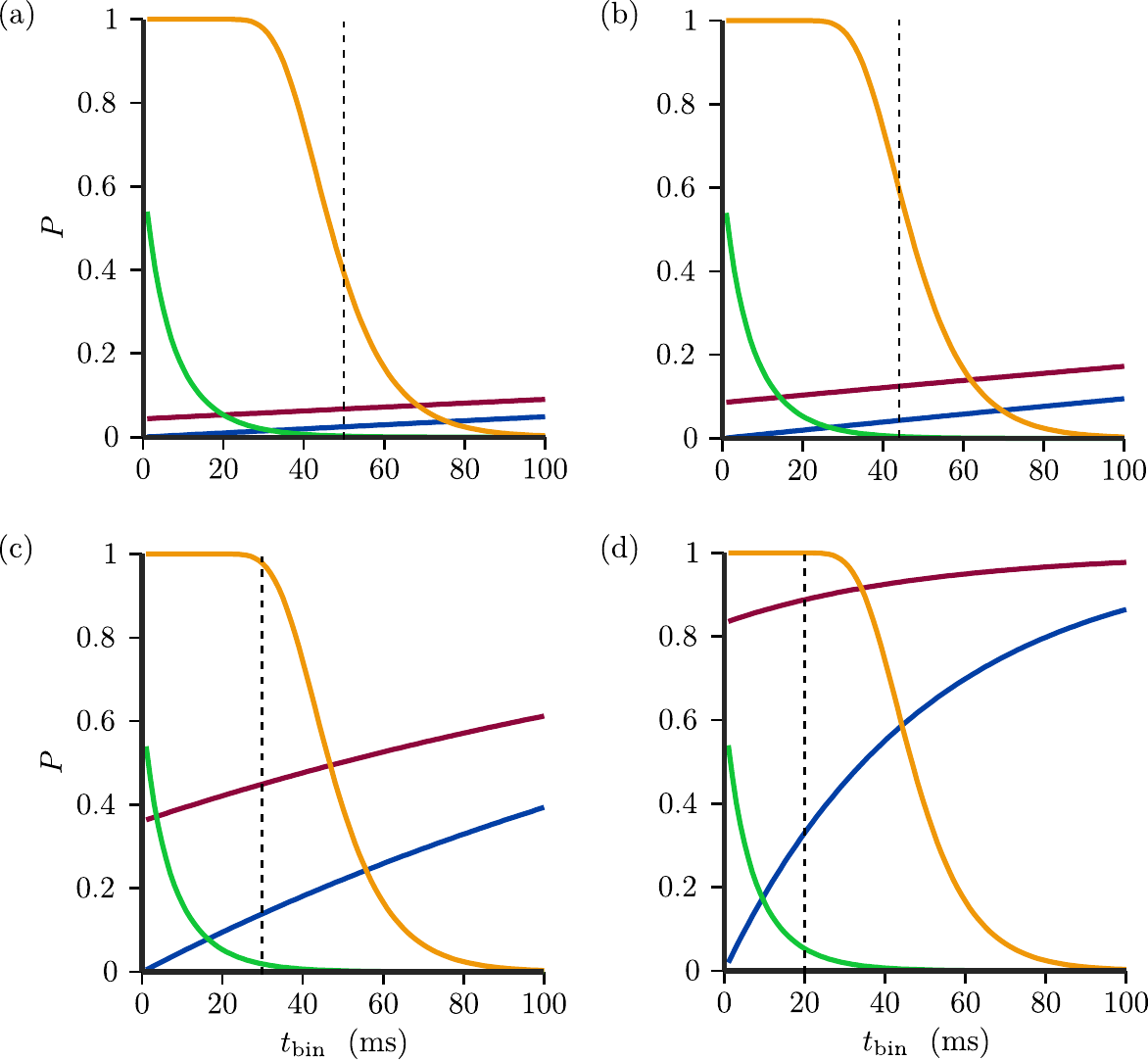}
\par\end{centering}
\caption{Probability estimates for joining and splitting of avalanches for
networks of (a) $N=50$, (b) $N=100$, (c) $N=500$, (d) $N=2000$
neurons and $\tau=10\,\text{ms}$, $f_{0}=0.01\,\text{Hz}$, $f_{\text{sat}}=2\,\text{Hz}$.
Each panel displays our estimates for joining and splitting first
spikes and average avalanches $P(\text{join first})$ Eq.~(\ref{bin:Pjoinfirst})
(blue), $P(\text{split first})$ Eq.~(\ref{eq:Psplitfirst3}) (green),
$P(\text{join average})$ Eq.~(\ref{bin:Pjoinaverage}) (magenta),
and $P(\text{split average})$ Eq.~(\ref{eq:Psplitaverage}) (orange)
versus bin size. The dashed line in (b) indicates the bin size chosen
for the data analysis in Figs.~3, 4, \ref{fig:FigB}, \ref{fig:FigG},
and \ref{fig:SubpopSpontA}; the other dashed lines indicate suitable
bin sizes for other network sizes.\label{fig:Binning}}
\end{figure}

We choose the bin size such that it keeps analytical probability estimates
for different errors that are generated by binning moderate, Fig.~\ref{fig:Binning}.
The considered errors are: (i) joining the initial spike of an avalanche
to the next avalanche, (ii) splitting the first spikes of the same
avalanche, (iii) joining an average size avalanche to the next one,
and (iv) splitting an average size avalanche. The resulting bin size
$t_{\text{bin}}$ depends on $f_{0},\tau$, $N$, and $f_{\text{sat}}$,
which are experimentally accessible from single neuron measurements,
anatomical data, and averaged spiking activity.

\begin{figure}
\begin{centering}
\includegraphics[width=0.5\columnwidth]{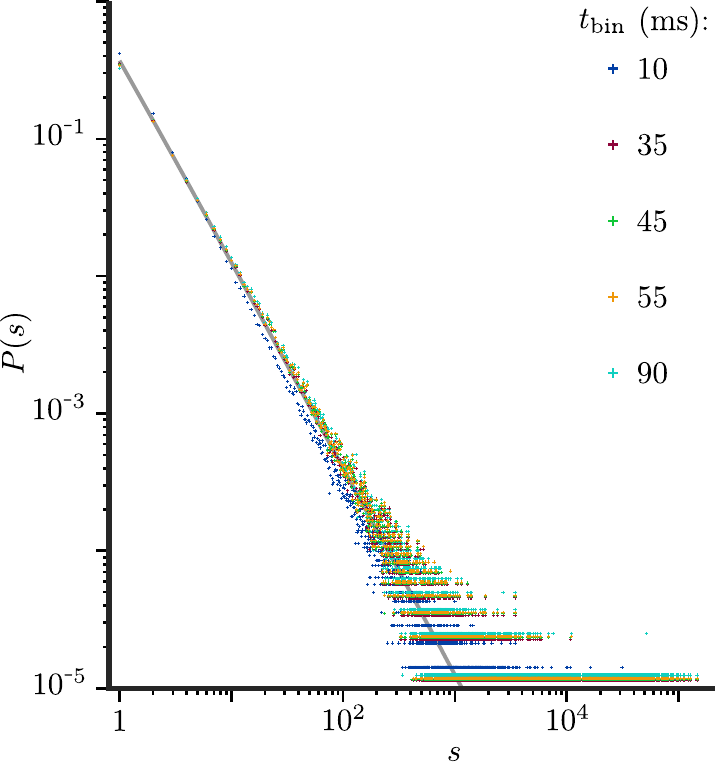}
\par\end{centering}
\caption{Robustness against large changes in bin size. The figure displays
avalanche size distributions obtained using bin sizes $t_{\text{bin}}=10\,\text{ms}$
(blue), $t_{\text{bin}}=35\,\text{ms}$ (red), $t_{\text{bin}}=45\,\text{ms}$
(green), $t_{\text{bin}}=55\,\text{ms}$ (orange), and $t_{\text{bin}}=90\,\text{ms}$
(cyan). The distributions obtained with $t_{\text{bin}}=35\,\text{ms}$,
$t_{\text{bin}}=45\,\text{ms}$, $t_{\text{bin}}=55\,\text{ms}$,
and $t_{\text{bin}}=90\,\text{ms}$ are similar, the one obtained
with $t_{\text{bin}}=10\,\text{ms}$ deviates, as expected from Fig.~\ref{fig:Binning}(b).
The analyzed spike data are generated by a network with our standard
parameters $N=100$, $\tau=10\,\text{ms}$, $f_{0}=0.01\,\text{Hz}$,
$f_{\text{sat}}=2\,\text{Hz}$; we usually use $t_{\text{bin}}=45\,\text{ms}$
for such data, Fig.~\ref{fig:Binning}(b).\label{fig:RobustnessBinning}}
\end{figure}

We first compute a simple estimate for the probability that binning
joins the initial spontaneous progenitor spike of an avalanche to
the next avalanche. Thereafter, we compute an estimate for the probability
of splitting an avalanche between its first two spikes. These two
probabilities yield a lower estimate for the probabilities of splitting
and joining avalanches, as avalanches extending beyond their initial
or first two spikes have higher probabilities of being joined to the
next or being split. Keeping the obtained probabilities small provides
an indication for an appropriate bin size, in particular because small
avalanches are frequent. To compute the probability of joining the
first spike of an avalanche to the next avalanche, we use that the
rate of spontaneous progenitor spikes in the network is $Nf_{0}$.
The interspike-interval (ISI) distribution of spontaneous spikes is
therefore $p_{\text{ISI}}(t)=Nf_{0}e^{-Nf_{0}t}$. The probability
of joining the initial spike of an avalanche to the following initial
spike of an avalanche is approximately (the bin will usually not start
at the first avalanche's start) the probability that the ISI between
progenitor spikes is less than $t_{\text{bin}}$,
\begin{align}
P(\text{join first}) & \approx\int_{0}^{t_{\text{bin}}}p_{\text{ISI}}(t)dt=1-e^{-Nf_{0}t_{\text{bin}}}.\label{bin:Pjoinfirst}
\end{align}
We now estimate the probability of splitting the first two spikes
of an avalanche. This is approximately the probability that the second
spike of the avalanche will occur more than $t_{\text{bin}}$ apart
from the first, $P(\text{split first})\approx P(t_{\text{bin}}<t_{2}<\infty)$,
where $t_{2}$ is time of the second spike. The first spike increases
the firing rate of the system by $\sigma/\tau$, so $P(\text{split first})$
can be written as
\begin{align}
P(\text{split first}) & \approx P(t_{\text{bin}}<t_{2}<\infty)=P(t_{\text{bin}}<t_{2})-P(t_{2}=\infty)\label{eq:Psplitfirst1}\\
 & =e^{-\int_{0}^{t_{\text{bin}}}\frac{\sigma}{\tau}e^{-t/\tau}dt}-e^{-\int_{0}^{\infty}\frac{\sigma}{\tau}e^{-t/\tau}dt}\label{eq:Psplitfirst2}\\
 & =e^{-\sigma(1-e^{t_{\text{bin}}/\tau})}-e^{-\sigma}.\label{eq:Psplitfirst3}
\end{align}
To assess how to choose the bin size to keep probabilities of joining
and splitting larger avalanches moderate as well, we derive similar
estimates for avalanches with average length and duration. The probability
of joining an avalanche of average duration $\bar{T}$ to the next
one is approximately
\begin{align}
P(\text{join average}) & \approx\int_{0}^{\bar{T}+t_{\text{bin}}}p_{\text{ISI}}(t)dt=1-e^{-Nf_{0}(\bar{T}+t_{\text{bin}})};\label{bin:Pjoinaverage}
\end{align}
we compute $\bar{T}$ by numerically integrating $\bar{T}=\int_{0^{-}}^{\infty}Tp(T)dT$,
where $p(T)$ is given by Eq.~(10). To estimate the probability of
splitting an avalanche of approximately average size $\bar{s}=1/(1-\sigma)$
{[}mean of the Borel distributed avalanche size, Eq.~(4){]}, we first
note that the split may occur about $\bar{s}-1=\sigma/(1-\sigma)$
times, which is the average number of offspring spikes in the branching
process. We assume that the excitation from the previous avalanche
spike has decayed to nearly zero when another one occurs, such that
the next spike is generated by the previous only. Each spike of the
avalanche then increases the collective firing rate to about $\sigma/\tau$
above the level of spontaneous spiking, like a progenitor spike. Since
this implies that interspike-intervals are long, our assumption is
conservative and gives us a higher probability to split the avalanche.
It allows us to employ the already derived $P(\text{split first})$
as an estimate for splitting one of the $\sigma/(1-\sigma)$ intervals
between avalanche spikes. The probability of not splitting an average
size avalanche is approximately $(1-P(\text{split first}))^{\sigma/(1-\sigma)}$
and thus

\begin{equation}
P(\text{split average})\approx1-(1-P(\text{split first}))^{\frac{\sigma}{1-\sigma}}\approx1-\left(1-e^{-\sigma(1-e^{t_{\text{bin}}/\tau})}-e^{-\sigma}\right)^{\frac{\sigma}{1-\sigma}}.\label{eq:Psplitaverage}
\end{equation}
Replacing $P(\text{split first})$ by the probability of splitting
an avalanche of two spikes yields similar results.

Figure~\ref{fig:Binning} displays the four probabilities Eqs.~(\ref{bin:Pjoinfirst}),
(\ref{eq:Psplitfirst3}), (\ref{bin:Pjoinaverage}), and (\ref{eq:Psplitaverage})
against bin size. For small bin size there is a high probability of
splitting an avalanche, which would result in overestimating the decay
of avalanche distributions and possible critical exponents. The probability
of splitting an avalanche becomes negligible for large bin size. For
large bin size, however, there is a high probability of joining avalanches,
which would result in underestimating the decay of avalanche distributions
and possible critical exponents. Our above estimates allow us to choose
a bin size in between. For a faithful detection of the avalanche characteristics,
it is more important to avoid joining small avalanches and splitting
initial avalanche spikes, since they are most frequent. The bin size
should thus be chosen such that keeping the probabilities $P(\text{join first})$
and $P(\text{split first})$ small is attributed a higher weight than
keeping the probabilities $P(\text{join average})$ and $P(\text{split average})$
small. For the binning of our numerical data, we thus choose a bin
size in the middle of the interval delimited by the crossings of $P(\text{join first})$
and $P(\text{split first})$ on the left and $P(\text{join average})$
and $P(\text{split average})$ on the right.

Our results are not sensitive to the chosen bin size. Since the time
scales of avalanche dynamics and avalanche generation are well separated,
most avalanches are relatively short and far apart, so the probability
to join two avalanches increases slowly with bin size and there is
a large range of suitable ones, see Fig.~\ref{fig:RobustnessBinning}.

\providecommand{\noopsort}[1]{}\providecommand{\singleletter}[1]{#1}%